\pdfoutput=1

\documentclass[aps,twocolumn,amsmath,amssymb,preprintnumbers,floatfix,prb,superscriptaddress,longbibliography]{revtex4-2}

\usepackage[utf8]{inputenc}
\usepackage{newtxtext}
\usepackage[upint]{newtxmath}
\usepackage{microtype}
\usepackage{textcomp}
\usepackage{eucal}
\usepackage{bm}
\usepackage{siunitx}
\usepackage{comment}
\usepackage{lipsum}

\usepackage{enumerate}
\usepackage{amsfonts}
\usepackage{amsmath}
\usepackage{amssymb}
\usepackage{color}
\usepackage{soul}

\usepackage{graphicx}

\usepackage[colorlinks,allcolors=blue]{hyperref}
\usepackage[capitalize]{cleveref}

\definecolor{DarkRed}{rgb}{0.65,0,0}%
\definecolor{Green}{rgb}{0,0.3,0.3}
\definecolor{Purple}{rgb}{0.3,0,0.65}
\definecolor{Red}{rgb}{1,0,0}
\definecolor{Blue}{rgb}{0,0,0.85}
\definecolor{Magenta}{rgb}{1,0,1}

\newcommand{\be}{\begin{equation}}
\newcommand{\ee}{\end{equation}}

\newcommand{\prlsection}[1]{\textit{#1}.\kern0.05em---\kern0.05em\ignorespaces}

\begin{document}
\title{Superconducting phase diagram and spin diode effect via spin accumulation}
\author{Johanne Bratland Tjernshaugen}
\affiliation{Center for Quantum Spintronics, Department of Physics, Norwegian \\ University of Science and Technology, NO-7491 Trondheim, Norway}
\author{Morten Amundsen}
\affiliation{Center for Quantum Spintronics, Department of Physics, Norwegian \\ University of Science and Technology, NO-7491 Trondheim, Norway}
\affiliation{Nordita, KTH Royal Institute of Technology and Stockholm University, Hannes Alfvéns väg 12, SE-106 91 Stockholm, Sweden}
\author{Jacob Linder}
\affiliation{Center for Quantum Spintronics, Department of Physics, Norwegian \\ University of Science and Technology, NO-7491 Trondheim, Norway}

\begin{abstract}
Spin-split superconductors offer new functionality compared to conventional superconductors such as diode-effects and efficient thermoelectricity. The superconducting state can nevertheless only withstand a small amount of spin-splitting. 
Here, we self-consistently determine the spin transport properties and the phase diagram of a spin-split superconductor in the presence of an injected spin accumulation. 
Energy and spin relaxation are accounted for in the relaxation time approximation via a single effective inelastic scattering parameter. 
We find that the spin-splitting field in the superconductor enables a spin diode effect. 
Moreover, we consider the superconducting phase diagram of a system in contact with a spin accumulation and in the presence of spin relaxation, and find that
the inclusion of energy and spin relaxation alters the phase diagram qualitatively. In particular, these mechanisms turn out to induce a superconducting state in large parts of the phase diagram where a normal state would otherwise be the ground-state. We identify an FFLO-like state even in the presence of impurity scattering which can be controllably tuned on and off via the electrically induced spin accumulation. We explain the underlying physics from how the superconducting order parameter depends on the non-equilibrium modes in the system as well as the behavior of these modes in the presence of energy and spin relaxation when a spin-splitting field is present.

\end{abstract}
\maketitle

\section{Introduction} 
In the field of superconducting spintronics \cite{linder_nphys_15, eschrig_rpp_15}, the aim is to use superconductors to either enhance spintronics phenomena compared to their 
behavior in the normal state or to identify entirely new spin-dependent phenomena without any normal state counterpart. There exists a large body of literature on this topic 
\cite{bergeret_rmp_05, buzdin_rmp_05, beckmann_jpcm_16, bergeret_rmp_18, amundsen_arxiv_22, bobkova_jpcm_22, geng_sst_23}, which has the potential to lead to cryogenic applications in fields such 
as quantum sensing and information technology.

One particular topic that in recent years \cite{quay_nphys_13,  hubler_prl_12, jeon_natmat_18} has gained increasing experimental attention is that of spin injection into 
superconductors \cite{ yamashita_prb_02, takahashi_prb_03, morten_prb_04}. Key questions in this context include the spatial extent that a 
spin current penetrates into the superconducting region \cite{takahashi_prb_03, morten_prb_04, silaev_prl_15}, if a resistive spin current can be converted into a spin supercurrent 
carried by Cooper pairs \cite{jeon_natmat_18, ouassou_scirep_19,  ouassou_prb_17, montiel_prb_18, montiel_prb_23}, how large spin accumulation that the 
superconducting state can sustain before transitioning to the normal state \cite{bobkova_prb_11}, and if the spin-Hall \cite{dyakonov_jetp_71, dyakonov_pla_71, hirsch_prl_99} efficiency can be made much larger in superconductors 
compared to other materials \cite{wakamura_nmat_15, espedal_prb_17}.

A particularly interesting situation arises when superconductivity coexists with a spin-splitting field \cite{tedrow_prl_71, tedrow_prb_73, meservey_physrep_94}. Besides emergent 
triplet Cooper pairs with a non-local correlation in time \cite{berezinskii_jetp_74, linder_rmp_19}, such systems are interesting from a practical point of view since they exhibit large 
thermoelectric effects \cite{kalenkov_prl_12, machon_prl_13, ozaeta_prl_14, kolenda_prl_16} and efficient 
spin-charge conversion via the inverse spin Hall effect \cite{jeon_acsnano_21, kamralinder_arxiv_23}. Although the coexistence of superconductivity with an order parameter 
$\Delta$ and spin-splitting $m$ is limited by the 
Clogston-Chandrasekhar limit \cite{clogston_prl_62, chandrasekhar_apl_62} $m = \Delta_0/\sqrt{2}$ where $\Delta_0$ is the zero-temperature bulk superconducting gap, several works have lately found ways to circumvent this limit by driving the system out of equilibrium 
\cite{ouassou_prb_18}, or by considering multiband 
systems \cite{ghanbari_prb_22, salamone_prb_23}. Increasing the permissible spin-splitting in superconductors is an interesting prospect since it could further enhance thermoelectricity 
and spin-charge conversion in the 
superconducting state. 

\begin{figure}[t!]
\includegraphics[width=0.95\columnwidth]{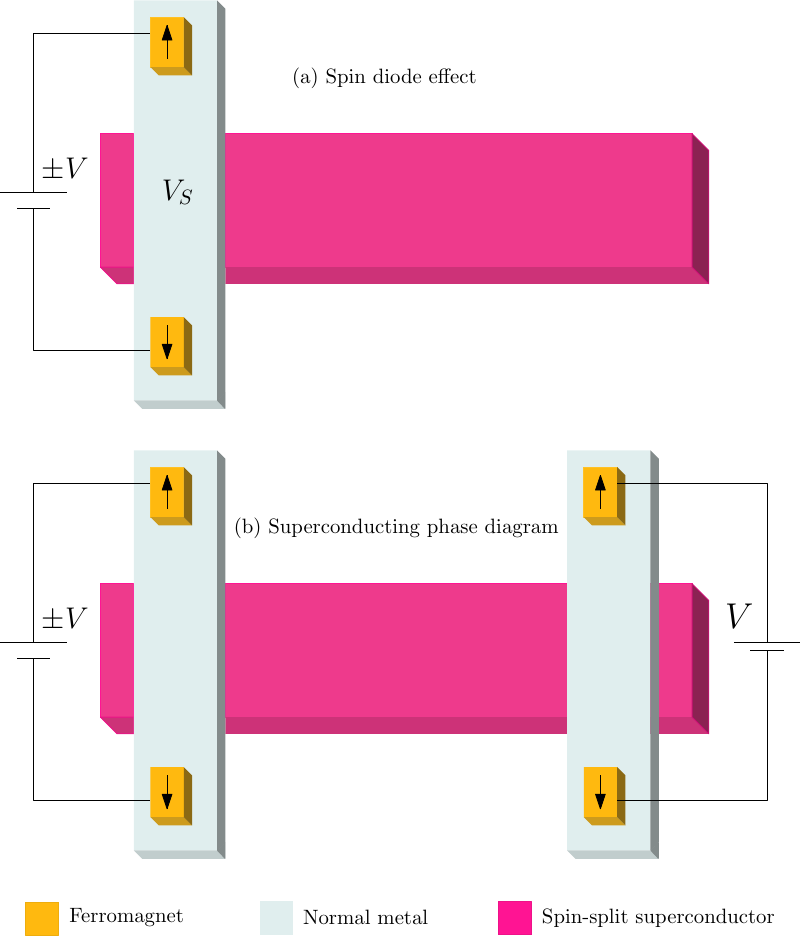}
	\caption{(Color online) Suggested experimental setups for the predicted effects. (a) Spin injection into a spin-split superconductor will be quantitatively different depending on the polarity of the injected spin, resulting in a spin-diode effect. (b) Interfacing a spin-split superconductor with two spin accumulations on each side that can be polarized parallel or oppositely strongly influences the superconducting phase diagram. The spin-split superconductor is achieved by taking a thin film superconductor and either applying an in-plane magnetic field or growing a ferromagnetic insulator under it. As for actual materials, the superconductor is a BCS superconductor such as Al, NbN or Pb. The normal metal could for example be Cu, and the ferromagnets could for example be Co, Fe or Ni. Examples of ferromagnetic insulators include EuS and GdN.}
	\label{fig:model}
\end{figure}

Motivated by several recent experimental advances \cite{hubler_prl_12, quay_nphys_13, jeon_natmat_18} with regard to spin injection in superconductors, we predict in this paper 
several effects that take place when a spin accumulation is injected into a spin-split superconductor. The spin-splitting is known to be realized experimentally either by proximity to a ferromagnetic insulator or via the Zeeman-splitting caused by an in-plane magnetic field. We first investigate the spin current $I_s$ arising in this setup [see Fig. \ref{fig:model}(a)]. 
We find that a spin diode effect occurs, such that the 
spin current satisfies $I_s(V_s) \neq -I_s(-V_s)$. The influence of spin and energy relaxation in the superconductor on this diode effect is determined within the relaxation time approximation. Next, motivated 
by the results of Ref. \cite{ouassou_prb_18} who found that a charge accumulation could stabilize the superconducting state for spin-splittings far above the Clogston-Chandrasekhar 
limit, we determine the superconducting phase diagram of a spin-split superconductor that is in contact with a spin accumulation. We consider a spin-valve setup where the injected spin 
accumulation from the reservoirs has the same (P) or opposite polarity (AP) arising in this setup [see Fig. \ref{fig:model}(b)]. In the P case, we find similar results as Ref. \cite{bobkova_prb_11}: the spin injection stabilizes the 
superconducting state as it completely removes the pair-breaking effect of the spin-splitting. In the AP state, a Fulde-Ferrel-Larkin-Ovchinnikkov (FFLO)-like \cite{fulde_pr_64, larkin_jetp_65} state appears and the phase diagram is 
qualitatively different from the P case. In both configurations, we determine the effect of spin and energy relaxation in the system, captured by a single parameter $\delta$ modelling inelastic scattering in the relaxation time approximation. Interestingly, we find that $\delta$ 
can substantially enhance or diminish the superconducting part of the phase diagram depending on the polarity of the spin accumulation. We explain this effect in terms of how the 
non-equilibrium energy and spin distribution functions in the superconductor are affected by $\delta$, and how the self-consistency equation for the superconducting order parameter in turn depends on these distribution functions. Our results on the stability of a spin-split superconducting state with respect to 
spin injection, and in particular the role of inelastic scattering, are relevant with respect to recent works \cite{jeon_acsnano_21, kamralinder_arxiv_23} that have studied, 
experimentally and theoretically, the inverse spin Hall effects in spin-split superconductors.

\section{Theory} 
We will make use of non-equilibrium Keldysh Green function theory \cite{bergeret_rmp_18, Belzig1999} to compute the superconducting phase diagram and spin diode effect that occurs in the system. This is a well-established theoretical framework that is convenient to use in the experimentally relevant limit of diffusive transport. It is also known to compare well, even quantitatively, with experiments performed on hybrid structures involving mesoscopic superconductors. 

In the quasiclassical Keldysh framework, the Green function is a momentum averaged $8\times 8$ matrix in Keldysh $\otimes$ Nambu $\otimes$ spin space,
\begin{equation}
    \check{g} = \begin{pmatrix}
        \hat{g}^R & \hat{g}^K \\ 0 & \hat{g}^A
    \end{pmatrix}.
\end{equation}
The retarded Green function $\hat{g}^R$ and the advanced Green function $\hat{g}^A$ are related by $\hat{g}^A = - \hat{\rho}_4 (\hat{g}^R)^{\dagger}\hat{\rho}_4$. The matrices $\hat{\rho}_n$ are defined as $\hat{\rho}_n = \tau_0 \otimes \sigma_n$ for $n\in \{0,1,2,3\}$ and $\hat{\rho}_n = \tau_3 \otimes \sigma_{n-4}$ for $n \in \{4,5,6,7\}$. The Keldysh Green function is 
\begin{align}
\hat{g}^K = \hat{g}^R\hat{h}- \hat{h}\hat{g}^A,
\end{align}
where $\hat{h} = h_m \hat{\rho}_m$ is a block diagonal matrix termed the distribution function. The Green function obeys the Usadel equation \cite{usadel_prl_70},
\begin{equation}
    \xi^2\partial_x(\check{g}\partial_x\check{g}) = -i[E\hat{\rho}_4+\hat{\Delta}+\hat{M}+\check{\Sigma}_\text{inelastic},\check{g}]/\Delta_0.
\end{equation}
Here, $\xi$ is the correlation length and $E$ is the quasiparticle energy. Superconductivity is accounted for by $\hat{\Delta} = \text{antidiag}(\Delta, -\Delta, \Delta^*, -\Delta^*)$, where $\Delta$ is the superconducting order parameter. A spin splitting field in the $z$-direction with strength $m$ enters the equation as $\hat{M} = \text{diag}(m\sigma_z, m \sigma_z)$. The inelastic scattering is accounted for via a self-energy:
\begin{align}
\check{\Sigma}_\text{inelastic} = \begin{pmatrix}
    i\delta\hat{\rho}_4 & 2i\delta h_\text{eq} \hat{\rho}_4 \\
    0 & -i\delta\hat{\rho}_4  \\
\end{pmatrix}
\end{align}
where $\delta$ quantifies inverse relaxation time and $h_\text{eq} = \tanh(\beta E/2)$. Formally, such a term can be derived by considering a tunneling coupling between the material and an effective thermal bath (reservoir) which is at equilibrium with a tunneling coupling that is independent on energy and momentum. This is the relaxation time approximation. It causes all non-equilibrium modes to decay with the same rate, forcing the distribution function matrix $\hat{h}$ in our setup to decay toward the equilibrium value $h_\text{eq}\hat{1}$ of the bath. Such a decay can be used as a simple model for actual microscopic processes that break energy conservation and spin conservation, such as electron-phonon scattering and magnetic impurity scattering, respectively. We note that this approach is not accurate when studying systems with charge currents, as it would not be conserved within this model. In the present system, on the other hand, there exists no charge current or charge mode, and so the relaxation time approximation employed here does not lead to unphysical results. From here on forward, inelastic scattering thus refers to both a decay of the energy mode $h_0$ and the spin mode $h_7$ in the system. We also note that quite different values for the spin relaxation length have been reported in Al \cite{poli2008spin} and NbN \cite{wakamura_prl_14, wakamura_nmat_15}, respectively. This means that by choosing different superconducting materials, one can effectively vary the strength of $\delta$ causing spin relaxation.

The superconductor is coupled to the reservoirs through tunneling contacts. Such couplings can be modelled by 
the Kupriyanov-Lukichev boundary conditions \cite{kupriyanov_zetf_88}, 
\begin{equation}
    \check{g}\partial_x\check{g} = \frac{\pm 1}{2L\zeta}[\underline{\check{g}},\check{g}],
\end{equation}
where $L$ is the length of the superconductor, $\zeta$ is an interface parameter and $\underline{\check{g}}$ denotes the Green function in the reservoir. The sign $+1$ is used at the interface where $x=0$, while $-1$ is used when $x=L$. The distribution function for a reservoir with a spin accumulation $V_s$ is 
\begin{equation}
    \underline{\hat{h}} = h_+(\hat{\rho}_0+\hat{\rho}_7)/2 + h_-(\hat{\rho}_0-\hat{\rho}_7)/2
\end{equation}
 with $h_{\pm} = \tanh{[(E \pm |e|V_s)/2T]}$. The spin current $I_s = I_s(x/\xi)$, given by
\begin{equation}
    I_s = I_0 \int_{0}^{\infty} \text{Re Tr}\left[\hat{\rho}_7\left(\check{g} \frac{\partial \check{g}}{\partial (x/\xi)} \right)^K \right]\text{d}\left(\frac{E}{\Delta_0}\right),
\end{equation}
flows into the reservoir for $V_s>0$. Here, $I_0 = N_0 \Delta_0^{2}\xi A / 16 $ where $N_0$ is the normal-state density of states at the Fermi level and $A$ is the interfacial contact area. When inelastic scattering is present, the spin current is not conserved due to spin relaxation. Hence, the spin current depends on position $x$. The spin current also depends on the value of the spin accumulation, and we recall that a diode effect is defined by $I_s(V_s) \neq - I_s(-V_s)$. We define the upper and lower critical spin accumulations $V_s^{+}$ and $V_s^{-}$ by the spatial average $I_s(V_s^{\pm}) = \pm 0.01 I_0$, respectively. The efficiency $\eta$ of the spin diode is given by
\begin{equation}
    \eta = \frac{|V_s^{+}|-|V_s^{-}|}{|V_s^{+}|+|V_s^{-}|}.
\end{equation}
In the absence of a diode effect, $\eta =0$.

We have considered a superconductor of length $L=8\xi$, and reservoirs with temperature $T=0.01T_c$. The interface parameter was set to $\zeta=3$ corresponding to a high barrier resistance compared to the normal state resistance. 

The retarded Usadel equation was Riccati parametrized \cite{schopohl_prb_95, konstandin_prb_05}, and the kinetic equations for the distribution function were parametrized as in Ref. \cite{ouassou_prb_18}. Numerically, these equations are solved by using fixed point iterations: first, we guess a value for the superconducting order parameter, and then we solve the retarded and kinetic equations successively until the change in the order parameter falls below some tolerance. 
The order parameter in a system with only the energy mode $h_0$ and the spin mode $h_7$ being nonzero is given by the gap equation
\begin{equation}
    \Delta = \frac{N_0\lambda}{4}\int_{-\omega_c}^{\omega_c}  [\hat{g}^R_{23}(E) + \hat{g}^R_{14}(-E)][h_0(E)-h_7(E)]\text{d} E.
\end{equation}
The coupling constant $\lambda$ is related through the cutoff energy by $\omega_c = \Delta_0\cosh{(1/N_0\lambda)}$. For some parameter sets, more than one numerically stable solution to the gap equation exist. We find different solutions by solving the Usadel equations with two different initial guesses on the order parameter: $\Delta = 0.01 \Delta_0$ (lower branch) and $\Delta = \Delta_0$ (upper branch). Superconductivity is defined by both branches converging to a nonzero value, and the normal state is defined by both branches converging to zero. Bistability is defined by the lower branch converging to zero and the upper branch converging to a nonzero value. In practice, convergence to zero must be defined by the gap having converged to a value less than some threshold value which we set to $0.005\Delta_0$. Physically, the states in the bistable region are either stable or metastable, but in the Keldysh framework it is hard to calculate free energies and thus determine which one is the physical ground state of the system. This also applies in parts of the superconducting region where the upper and lower branches converge to different, nonzero values. Finally, we note that in the absence of inelastic scattering ($\delta=0$), we model inelastic scattering in the retarded equations only using the Dynes approximation $E \rightarrow E + 0.01i\Delta_0$.

\section{Results}

\subsection{Spin diode effect}

\begin{figure}
    \centering
    \includegraphics{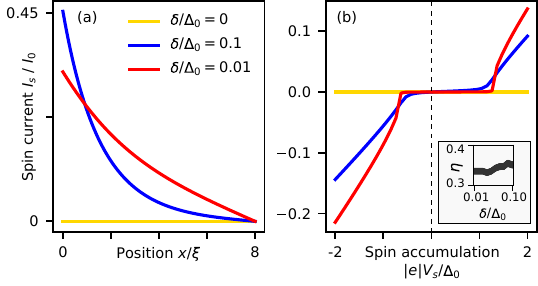}
    \caption{(Color online) Spin currents in setup A for different inelastic scattering parameters $\delta/\Delta_0$. The spin-splitting field in the superconductor is set to $m=0.32\Delta_0$. (a) Spatial profile of the spin current for spin accumulation $|e|V_s/\Delta_0 = 2$. (b) Spatially averaged spin current as a function of spin accumulation. The current is antisymmetric around $|e|V_s =0.32\Delta_0$. The inset shows the dependence of the diode efficiency $\eta$ on the inelastic scattering parameter. The numerical method for determining $|e|V_s^{\pm}/\Delta_0$ gave a small uncertainty $\mp0.01$ in determining these critical spin accumulations. The thickness of the line showing $\eta(\delta)$ in the inset accounts for this uncertainty in $V_s^{\pm}$. The uncertainty is thus much smaller than the actual magnitude of $\eta(\delta)$.
    } 
    \label{fig:diodeeffect-in-NSV}
\end{figure}

\begin{figure}
    \centering
    \includegraphics{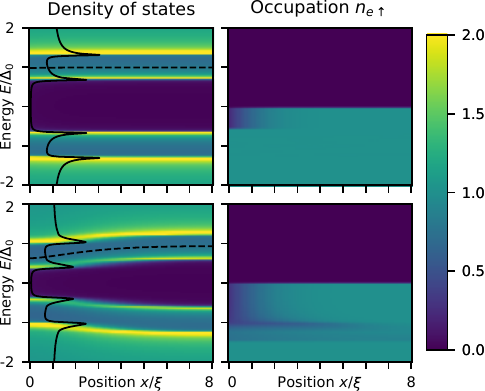}
    \caption{(Color online) Density of states and occupation number $n_{e\uparrow}$ for spin-$\uparrow$ electrons in setup A for $|e|V_s/\Delta_0 = 0.52$ (upper plots) and $|e|V_s/\Delta_0 = 1.47$ (lower plots). The dashed line shows the spatial profile of the gap $|\Delta|/\Delta_0$. The solid line shows the density of states at $x/\xi = 0$. The inelastic scattering parameter is set to $\delta/\Delta_0 = 0.01$.  }
    \label{fig:dos-nsv}
\end{figure}

\begin{figure*}
    \centering
    \includegraphics[width=2.1\columnwidth]{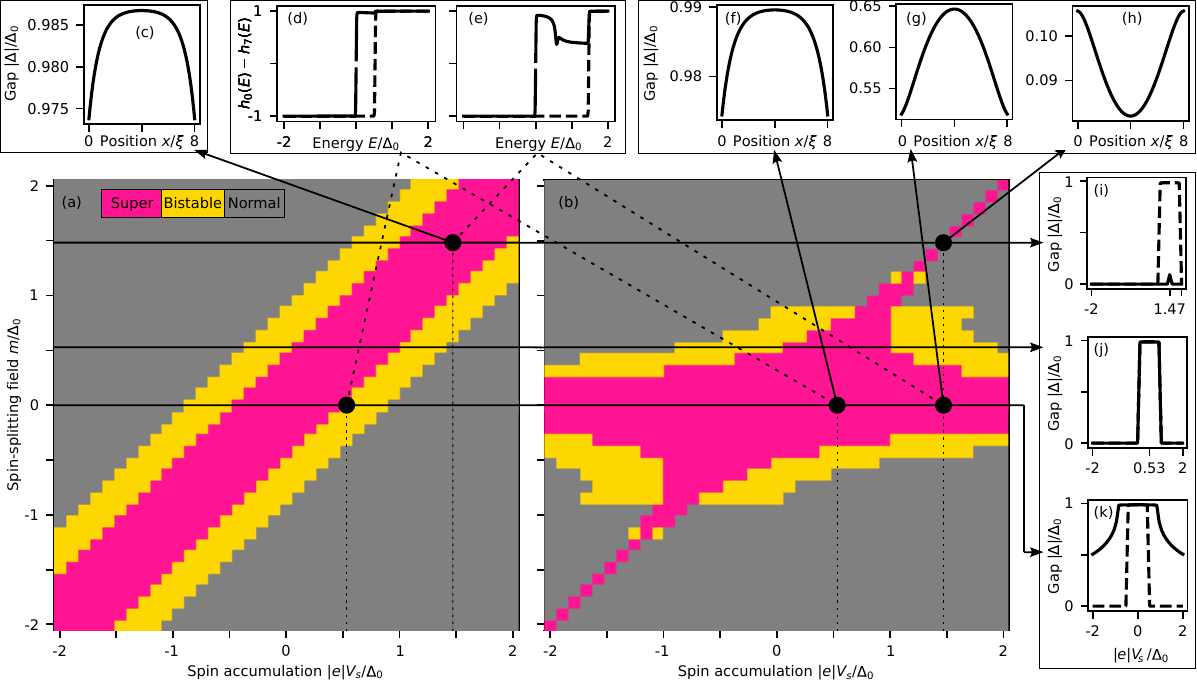}
    \caption{(Color online) Phase diagram for setup B with parallel spin accumulations showing the regions of superconductivity, bistability and the normal state for different spin accumulations and spin-splitting fields. In subfigure (a), inelastic scattering is not included in the kinetic equations, while in (b) it is included. The spatial profile of the superconducting gap for given parameter sets is given in subfigure (c) and subfigures (f-h). Subfigures (d-e) show the contribution from the distribution function in the gap equation in the middle of the superconductor for given parameter sets, and the spatially averaged superconducting gap for given spin-splitting fields is shown in subfigures (i-k). In subfigures (d-e) and (i-k), the dashed lines originate from the lower branch in phase diagram (a), while the solid lines originate from the lower branch in phase diagram (b). 
    }
    \label{fig:same-megafig}
\end{figure*}

In setup A, a spin current is injected from the reservoir with a spin accumulation $V_s$. This spin current decays in the superconductor when inelastic scattering is present, since this causes spin relaxation (which is modeled by a decay of the spin mode $h_7$). The spatial profile of the spin current is shown in Fig. \ref{fig:diodeeffect-in-NSV}(a). In setups with a vacuum interface, a non-zero spin current can only exist in the presence of inelastic scattering because the spin current must vanish at the vacuum interface. However, the spin current can exist even in the absence of inelastic scattering if vacuum is replaced by e.g. a normal metal in equilibrium. High values of the inelastic scattering parameter suppress the magnitude of the spin current because non-equilibrium modes decay faster when entering the superconductor. Fig. \ref{fig:diodeeffect-in-NSV}(b) shows the spatially averaged spin current at a spin splitting field $m/\Delta_0=0.32$ for different inelastic scattering parameters and spin accumulations. A spin diode effect is observed. 
The explanation for the diode effect lies in the spin-dependent particle-hole asymmetry present in the density of states (DOS) of the superconductor \cite{meservey_physrep_94}. Consider a positive spin accumulation $V_s>0$. The occupation number for spin-$\uparrow$ electrons is $n_{e\uparrow} = [1-(h_0+h_7)]/2$ which takes the form $f(E+|e|V_s)$ in the reservoir. This means that the spin-$\uparrow$ electron population occupies energies only up to the normal state Fermi level minus $|e|V_s$. A spin-$\uparrow$ current then starts to flow from the superconductor into the reservoir as soon the occupied energies in the reservoir are lowered by $|e|V_s = \Delta+m$. This is because the $\uparrow$ quasiparticles have a finite density of states inside the superconductor at $E=-\Delta-m$ and for increasingly negative energies. The same reasoning applies for the spin-$\downarrow$ electrons. These instead have a distribution function $n_{e\downarrow} = f(E-|e|V_s)$ inside the reservoir, so that they occupy energies up to the normal state plus $|e|V_s$. Spin-$\downarrow$ electrons will then start to tunnel into the superconductor as soon as there are available $\downarrow$-quasiparticle states. This happens when the spin-$\downarrow$ population has been increased by $|e|V_{s}=\Delta+m$, since the $\downarrow$-quasiparticle density of states in the superconductor becomes non-zero at $E=\Delta+m$ and higher energies. In this way, $\uparrow$ electrons flow from the S to the reservoir while $\downarrow$ electrons flow from the reservoir to S, both events contributing equally to the spin current and occurring when $|e|V_s$ exceeds $\Delta+m$. This is precisely where the current starts flowing in Fig. \ref{fig:diodeeffect-in-NSV}(b). The same type of analysis reveals that the current starts to flow at negative spin accumulations $|e|V_s = -\Delta+m$ or lower. 

As long as the inelastic scattering parameter is sufficiently small and superconductivity is present, it is reasonable to assume $V_s^{\pm} \approx m \pm \Delta$, which gives
\begin{equation}
    \eta =\begin{cases}
        \Delta/m, & m > \Delta \\
        m/\Delta, & m < \Delta.
    \end{cases}
\end{equation}
For small spin splitting fields $m<\Delta$, the efficiency of the diode increases linearly with $m$. When the spin splitting field increases, the efficiency is reduced for two reasons. Firstly, the value of the gap decreases. Secondly, the region with $I_s(V_s)=0$ is centered around $m$ and therefore large spin accumulations are required to observe the spin diode effect. The inset in Fig. \ref{fig:diodeeffect-in-NSV}(b) shows how the diode efficiency depends on the inelastic scattering parameter.

The spin relaxation causes the occupation number to change throughout the superconductor. This is shown in Fig. \ref{fig:dos-nsv}. The spatial gradient in the distribution function for quasiparticles inside the superconductor drives a spin current through the superconducting material itself. This spin current is, however, not conserved and dies out as one moves away from the reservoir interface. Moreover, we see that the suppression of the gap near the interface becomes considerable for large spin accumulations (lower left panel of Fig. \ref{fig:dos-nsv}).

This mechanism behind the diode effect is different from the one studied in Ref. \cite{amundsen_prb_22}, the reason being that we in the present study include the effect of spin and energy relaxation when determining the diode characteristics. When the injected spin accumulation combines with the spin-splitting to exceed the Clogston-Chandrasekhar limit, the superconductor transitions to the normal state. This was the underlying mechanism of the spin diode effect that has been studied previously \cite{amundsen_prb_22}. However, when spin and energy relaxation are included in the superconductor, the superconducting order parameter is able to recover sufficiently far away from the interface, so long that the spin-splitting field itself does not exceed the Clogston-Chandrasekhar limit. This means that the quasiparticle spin current flowing in the superconductor is dictated by the available density of states inside the superconductor, away from the interface, where the spin accumulation has decayed. This is in contrast to the mechanism \cite{amundsen_prb_22} where the spin accumulation induces a transition from the superconducting to the normal state. This highlights the importance of including relaxation mechanisms for non-equilibrium modes in the superconductor. The spin diode effect reported here arises due to an electron spin accumulation in the reservoir instead of a magnon spin accumulation studied in Ref. \cite{amundsen_prb_22}.

If we replace the vacuum interface in setup A with a normal metal in equilibrium ($|e|V_s=0$), we get a system where we can actually take advantage of the spin diode effect. The normal metal could for example be Cu. We explained the physics behind the spin diode effect in setup A by considering the normal metal/superconductor interface and by looking at the occupation and density of states in the normal and superconducting regions respectively. Swapping the vacuum interface with a normal interface will not change this argument. It will, however, allow a spin current to flow through the superconductor such that the spin diode can be part of a circuit. Contrary to setup A, a nonzero and conserved spin current will be allowed because the spin current does not vanish at the $x=8\xi$ interface.

\begin{figure*}
    \centering\includegraphics[width=2.1\columnwidth]{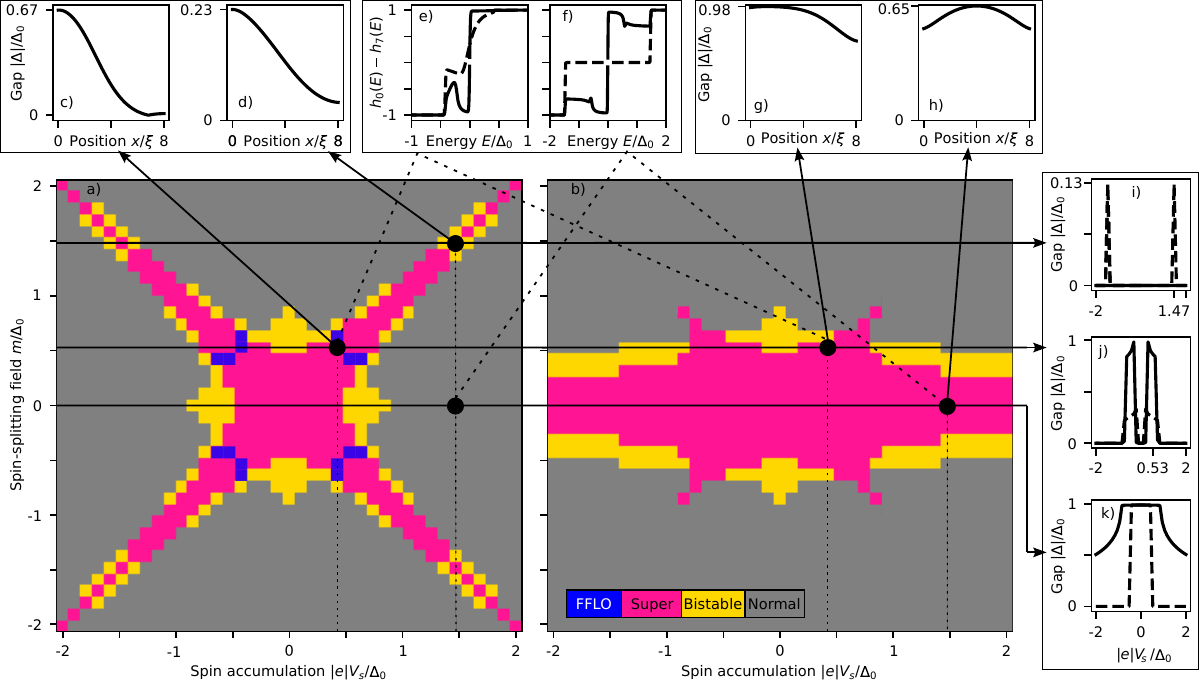}
    \caption{(Color online) Phase diagram for setup B with antiparallel spin accumulations showing the regions of superconductivity (including the FFLO-like state), bistability and the normal state for different spin accumulations and spin-splitting fields. In subfigure (a), inelastic scattering is not included in the kinetic equations, while in (b) it is included. The spatial profile of the superconducting gap for given parameter sets is given in subfigures (c-d) and (g-h). Subfigures (e-f) show the contribution from the distribution function in the gap equation in the middle of the superconductor for given parameter sets, and the spatially averaged superconducting gap for given spin-splitting fields is shown in subfigures (i-k). In subfigures (e-f) and (i-k), the dashed lines originate from the lower branch in phase diagram (a), while the solid lines originate from the lower branch in phase diagram (b). }
    \label{fig:oppo-megafig}
\end{figure*}

\subsection{Superconducting phase diagram}
We begin by considering the case where the same spin accumulation is applied to the ends of the superconductor [$+V$ on the left side of Fig. \ref{fig:model}(b)]. The phase diagram for this setup is shown in Fig. \ref{fig:same-megafig}. We here show that the phase diagram is qualitatively different depending on whether spin and energy relaxation are accounted for in the kinetic equations or not. For the case where inelastic scattering is not included in the kinetic equations, and thus only as a Dynes parameter \cite{dynes_prl_78} in the quasiparticle energies in the retarded part of the Green function, the results are shown in Fig. \ref{fig:same-megafig}(a). There exists a superconducting band in the phase diagram with width $\Delta_0$ following the line $m = |e|V_s$, and a region of width $0.5 \Delta_0$ outside this band is bistable.  The phase diagram is symmetric around $m=|e|V_s$, indicating that the combined effect of a spin-splitting field $m$ and a spin accumulation $|e|V_s$ amounts to an effective spin-splitting field $m_\text{eff}=m-|e|V_s$. This finding was first reported in Ref. \cite{bobkova_prb_11}. The spatially averaged value of the superconducting gap is shown as the dashed lines in Fig. \ref{fig:same-megafig}(i-k), and the spatial profile of the gap for one parameter set is shown in Fig. \ref{fig:same-megafig}(c) . It is seen that the value is either zero, corresponding to the normal state, or $\Delta_0$, which is the bulk gap value. This indicates that the system behaves like a zero-temperature, bulk superconductor with an effective spin-splitting field $m_\text{eff}$. The distribution function inside the superconductor is identical to the distribution function in the reservoirs [see the dashed lines in Fig. \ref{fig:same-megafig}(d-e)] because there is no inelastic scattering in the kinetic equations which changes the occupation numbers. The above reasoning can also be proven analytically. Inserting the general expression for the distribution function and the bulk expression for the retarded Green function into the gap equation gives 
\begin{equation}
\begin{split}
    \Delta &= \frac{N_0\lambda}{4}\int_{-\omega_c}^{\omega_c} \text{d} E \frac{-2\Delta \Theta((E-m)^2-\Delta^2)}{\sqrt{(E-m)^2-\Delta^2}} \tanh \left( \frac{E-|e|V_s}{2T} \right).
\end{split}
\end{equation}
Defining a new integration variable $E' = E - |e|V_s$ shows that the spin-splitting field and the spin accumulation show up only in the combination $m - |e|V_s$. This demonstrates that the effect of the spin accumulation and the spin-splitting field indeed is the equilibrium system with an effective field  $m - |e|V_s$. 

We now show that inelastic scattering in the kinetic equations changes the phase diagram qualitatively. The results are shown in Fig. \ref{fig:same-megafig}(b). Spatial profiles for the gap are shown in Fig. \ref{fig:same-megafig}(f-h), and the spatially averaged values of the gap are shown as solid lines in Fig. \ref{fig:same-megafig}(i-k). 
Firstly, the gap is suppressed for high spin-splitting fields $m$ because inelastic scattering destroys non-equilibrium modes as one moves away from the edges of the superconductor where the spin injection takes place. The part of the distribution function that enters the gap equation [see the solid lines in Fig. \ref{fig:same-megafig}(d-e)] does therefore no longer compensate for the spin-splitting field. However, at the edges of the superconductor the non-equilibrium modes are not suppressed and the compensating effect of the spin splitting field and the spin accumulation gives a non-zero gap as seen in Fig. \ref{fig:same-megafig}(h). Therefore, part of the system is superconducting as long as $m$ is sufficiently close to $|e|V_s$. The width of the superconducting stripe in Fig. \ref{fig:same-megafig}(b) decreases as $m$ and $|e|V_s$ grow, until at some value $m = |e|V_s$ the suppression of the gap in the bulk is so large that the average value of the gap falls below the threshold for superconductivity, and thus the stripe disappears. 
Secondly, superconductivity survives for high spin accumulations as long as the spin-splitting field is sufficiently small. This is because the superconductor behaves like a spin-split superconductor in equilibrium far from the edges. Due to superconductivity being suppressed at the edges of the superconductor, the spatially averaged superconducting gap decreases when the spin accumulation increases as seen in Fig. \ref{fig:same-megafig}(k). In effect, spin and energy relaxation \textit{induces} superconductivity in large regions of the phase diagram which would otherwise have a normal ground-state.

Next, we consider the case where oppositely polarized spin accumulation is applied to the ends of the spin-split superconductor [$-V$ on the left side of Fig. \ref{fig:model}(b)]. Reversing the sign of the spin accumulation in one of the reservoirs qualitatively changes the superconducting phase diagram, Fig. \ref{fig:oppo-megafig}. Firstly, it restores the symmetry of the phase diagram for positive and negative $V_s$ that was absent in Fig. \ref{fig:same-megafig}. This can be understood from a simple symmetry argument: physically rotating the entire system 180 degrees does not change the superconducting phase, but is equivalent to performing $V_s \to -V_s$ in both reservoirs. For the case without inelastic scattering in the kinetic equations, Fig. \ref{fig:oppo-megafig}(a), superconductivity is recovered for $m = |e|V_s$. This is understood by the cancellation of the effects of spin accumulation and the spin-splitting field at one interface, and the addition of the spin accumulation and the spin splitting field at the other interface. Spatial profiles of the superconducting gap are shown in Fig. \ref{fig:oppo-megafig}(c-d). We note that the phase diagram for $m>0, V_S>0$ now has a close resemblance to the phase diagram for an electric voltage-biased superconductor in a high field \cite{ouassou_prb_18}.

Interestingly, we find a stable Fulde-Ferrel-Larkin-Ovchinnikkov (FFLO)-like state characterized by the spatial inhomogeneity and sign change displayed by the order parameter $\Delta$. The gap is spatially asymmetric [see for instance Fig. \ref{fig:oppo-megafig}(d)]  since superconductivity is suppressed more at the interface where the spin-splitting and spin-accumulation combine destructively in the gap equation. However, in some regions of the phase diagram, the U(1) phase of the order parameter also changes from 0 to $\pi$ inside the superconductor [see for instance Fig. \ref{fig:oppo-megafig}(c)], providing an effective FFLO-like state. Such a state only appears in a limited part of the phase diagram Fig. \ref{fig:oppo-megafig}, similarly to its restricted regime in the equilibrium $m-T$ phase diagram \cite{matsuda_jpsj_07}. The appearance of the FFLO phase is caused by the inhomogeneity of the order parameter, which allows for the existence of $p$-wave triplet superconducting correlations even in the diffusive limit~\cite{eschrig_np_2008}. This is because such an inhomogeneity gives rise to a preferential scattering direction, so that momentum-dependent effects survive the momentum-averaging of highly diffusive materials. The simultaneous presence of both $s$- and $p$-wave superconducting correlations enables the system to mitigate the effect of spin-splitting, and is observable as a spatial modulation of the superconducting order parameter. Equivalent effects have also been observed in other non-equilibrium systems~\cite{moor_prb_2009,bobkova_prb_2013}. 
An interesting point here is that the FFLO-state can be obtained controllably by varying the spin accumulation and/or spin-splitting in the system, and even in the presence of frequent impurity scattering. In particular, the latter aspect usually prohibits a spontaneous FFLO state from appearing in spin-split superconductors, since it is highly sensitive to impurities under equilibrium conditions.  

The effect of inelastic scattering for antiparallel spin accumulations is prominent as seen in Fig. \ref{fig:oppo-megafig}(b). Superconductivity is no longer recovered for high $m = |e|V_s$ because the superconducting gap is suppressed both at one of the interfaces due to an effective spin splitting field $m_\text{eff} = m + |e|V_s$, and in the bulk of the superconductor due to the destruction of non-equilibrium modes. Spatial profiles of the gap are shown in Fig. \ref{fig:oppo-megafig}(g-h), and the destruction of the non-equilibrium modes are seen in Fig. \ref{fig:oppo-megafig}(e-f). For high spin accumulations and small spin splitting fields, superconductivity is recovered for the same reasons as for the setup with parallel spin accumulations. The spatially averaged gap for selected values of the spin splitting field is shown in Fig. \ref{fig:oppo-megafig}(i-k). We note in passing that the finding that inelastic scattering can assist in maintaining the superconducting state despite a large spin accumulation $V_{s}$ is relevant with regard to the recent proposal \cite{kamralinder_arxiv_23} of using spin injection into spin-split superconductors to obtain an inverse spin Hall effect that is orders of magnitude larger than in the normal state, as has been experimentally observed \cite{jeon_acsnano_20}.

We have considered a specific system with length $L=8\xi$ and interface parameter $\zeta=3$, but we can generalize the results to systems with different lengths and interface parameters. Increasing the length or decreasing the interface transparency
both cause  the bulk behavior of the superconductor to be more important. For systems with inelastic scattering, the non-equilibrium spin accumulation is present only in a small region close to the edge of the superconductor, and most of the superconductor resembles a spin-split superconductor in equilibrium. Systems without inelastic scattering are not sensitive to the length or the interface transparency, except the FFLO-like state which might be sensitive to the length. Decreasing the length or increasing the interface transparency enhance the effects of the non-equilibrium modes, causing the phase diagrams of Fig. \ref{fig:same-megafig}(b) and Fig. \ref{fig:oppo-megafig}(b) to mimic the phase diagrams without inelastic scattering given in Fig. \ref{fig:same-megafig}(a) and Fig. \ref{fig:oppo-megafig}(a).

\section{Conclusion} 
Motivated by the fact that spin-split superconductors show a high potential for cryogenic applications within thermoelectricity and non-reciprocal transport, we have self-consistently studied spin transport properties and the phase diagram of such a superconductor in the presence of an injected spin accumulation. Energy and spin relaxation are accounted for in the relaxation time approximation via a single effective inelastic scattering parameter. 
We find a quasiparticle spin-diode effect in the presence of a spin splitting field. 
The superconducting phase diagram changes qualitatively in the presence of the above-mentioned relaxation mechanisms. In particular, they can induce a superconducting state in large parts of the phase diagram where a normal state would otherwise be the ground-state. We also find a tunable FFLO-like state with a 0-$\pi$ transition of the order parameter that is robust toward impurity scattering. These results establish the conditions required to have a robust superconducting state coexisting with spin-splitting when the system is driven out of equilibrium by spin injection. This is important since actual applications drawing upon the interesting quantum physics of spin-split superconductors necessarily will have to involve non-equilibrium conditions, whether one deals with spin-dependent thermoelectricity or non-reciprocal transport.

 \begin{acknowledgments}
J. A. Ouassou and L. G. Johnsen are thanked for useful discussions. This work was supported by the Research
Council of Norway through Grant No. 323766 and its Centres
of Excellence funding scheme Grant No. 262633 “QuSpin.” Support from
Sigma2 - the National Infrastructure for High Performance
Computing and Data Storage in Norway, project NN9577K, is acknowledged.
 MA acknowledges support from the Swedish Research Council (Grant No. VR 2019-04735 of Vladimir Juri\v{c}i\'c). Nordita is supported in part by NordForsk.
 \end{acknowledgments}

\bibliography{refs}

\end{document}